\documentclass[showpacs,showkeys]{revtex4}

\usepackage{graphicx}
\usepackage{amssymb}
\usepackage{bm}

\renewcommand{\baselinestretch}{2}

\begin{document}
\setcounter{page}{1}
\newcommand{\re}[1]{(\ref{#1})}
\newcommand{\lab}[1]{\label{#1}}
\newcommand{\ci}[1]{\cite{#1}}
\renewcommand{\baselinestretch}{1.25}
\newcommand{\bfr}{\begin{flushright}}
\newcommand{\bfl}{\begin{flushleft}}
\newcommand{\efl}{\end{flushleft}}
\newcommand{\efr}{\end{flushright}}
\newcommand{\bc}{\begin{center}}
\newcommand{\ec}{\end{center}}
\newcommand{\be}{\begin{equation}}
\newcommand{\ee}{\end{equation}}
\newcommand{\bea}{\begin{eqnarray}}
\newcommand{\eea}{\end{eqnarray}}
\newcommand{\ba}{\begin{array}}
\newcommand{\ea}{\end{array}}
\newcommand{\edc}{\end{document}}
\newcommand{\ul}{\underline}
\newcommand{\ri}{\rightarrow\infty}
\newcommand{\li}{\leftarrow\infty}
\newcommand{\ra}{\rightarrow}
\newcommand{\la}{\leftarrow}
\newcommand{\ds}{\displaystyle}
\newcommand{\dsf}{\displaystyle\frac}
\newcommand{\dt}{\Delta{t}}
\newcommand{\il}{\int\limits}
\newcommand{\pal}{\partial}
\newcommand{\xxx}{{\it{X}}}
\newcommand{\bone}{{\bf 1}}
\newcommand{\gComment}[1]{}
\renewcommand{\gComment}[1]{\textcolor{red}{Gerardo: #1}}

\title[]{Examination of Current-Induced Magnetic Field in the Slab\\ Geometry: Possible Origin of Spin Hall Effect}
\author{B. \surname{Abdullaev}}
\email[E-mail: ]{babdullaev@nuuz.uzsci.net}
\author{M. \surname{Choi}}
\author{C.-H. \surname{Park}}
\affiliation {Research Center for Dielectric and Advanced Matter
Physics, Department of Physics, Pusan National University, Busan 609-735}

\date[]{Received 22 November 2006}

\begin{abstract}

We estimate the strength of current-induced magnetic field (CIMF) in
the two-dimensional slab geometry for Spin Hall Effect (SHE)
observed recently by  Kato {\it et al.} and Wunderlich {\it et al.} and show
that if the factor $gm^*/m$, where $g$ is the Lande factor and $m^*$ and
$m$ are effective and pure masses, respectively, is equal to
the numerical value at the surface of the semiconductor, then the CIMF can
describe the SHE.

\end{abstract}

\maketitle

\section{Introduction}

Recent observation of electron  Spin Hall Effect (SHE) by Kato {\it et
al.} \ci{kato} and hole SHE by Wunderlich {\it et al.}  \ci{wunderlich} has
attracted much attention from condensed-matter physicists, since
spin polarization on nonmagnetic semiconductor thin-film edges
has been induced by longitudinal electric current. Spins \ci{kato}
are polarized along the direction perpendicular to the current. However,
Kato {\it et al.} claimed that the current-induced  spin-polarization
should be below the experimental capability to detect it.

\section{Theory and Mathematical Treatment}

For a theoretical explanation of the SHE,  theories based on the
spin-orbit (SO) interaction have been widely developed. Although the
SO interaction is relativistic \ci{landau} of second order on ratio
of electron velocity to light velocity, it is supposed \ci{rashba}
that a band splitting induced by the SO interaction is of the same
scale as the energy gap between the conduction and valence bands.
Two kinds of microscopic scenarios employing the SO interaction have
been investigated: extrinsic \ci{dyakonovandothers} as a result of
asymmetric scattering for up and down spins, and intrinsic
\ci{murakamiandothers} connected only with the band structure of
semiconductors. These approaches have been extensively discussed
\ci{huandothers} in the literature. Nomura {\it et al.} \ci{nomura}
combined the theoretical calculations for the intrinsic effect with
the experimental data. However, among these complicated treatments,
it seems that the role of the Zeeman splitting of the electronic
energies by the current-induced magnetic field (CIMF) for the SHE has
so far not been sufficiently investigated. Kato {\it et al.} considered that the CIMF
strength is not large enough to explain the observed SHE. In this
paper, we show that, by taking into account some material
parameters, the CIMF can be responsible for the SHE and the
spin polarization of the electron gas in semiconductors. Namely, we will
show that the estimated CIMF on edges of slab geometry of samples
and the numerical value of the factor $gm^*/m$ on the surface of
the semiconductor can provide a sufficient value required for SHE. More
explicitly, we show that the component of the CIMF perpendicular to
the plane of a slab around the edge is divergent as the logarithm of the ratio
of the width to the thickness of the slab. It is zero at the middle
of the width. Therefore, in the case of infinitesimal thickness,
the spins might be polarized perpendicular to the semiconductor
mainly around the edges. This structure of the magnetic field might
provide the asymmetric Hall Effect, when both edges of the sample are
charged with same-sign charges (we have called the effect
asymmetric because in the conventional symmetric Hall Effect the two edges
are charged with opposite-sign charges). Additionally, we will
repeat the calculation by Kato {\it et al.} to estimate the strength of CIMF
and confirm that it equals ours. At the condition $gm^*/m
\rightarrow 2$, {\it i.e.}, when the factor $gm^*/m$ tends to have a
vacuum value on the surface of the semiconductor (we can assume that the
surface-monolayer non-magnetic atoms are to be in vacuum and put for
their electrons $m^*\approx m$ and  $g= 2$ \ci{landau3}), the
numerical value of CIMF strength will be enough to create
experimental SHE.

The absolute value of magnetic field $\vec H$ for the infinite
length cylindrical-geometry metallic sample, as a solution of the
Maxwell equation, $\vec \nabla \times H = (4 \pi /c)\vec j$, written
in the integral form, $\oint \vec H d \vec l =(4 \pi /c)\int \vec j
d \vec f$, has the form $ H = 2 J(r)/(cr)$, if the density of the
current $j(r)$  is constant at a fixed radial distance $r$ from
the longitudinal axis of the conductor. Here, magnetic field $\vec H$ is
along the closed curve $\vec l$ with radius $r$, taken around the
conductor on a transverse section to it, and $J(r)$ is the
current flowing through the area $\pi r^2$.

The experimental film slab can be modeled as a finite number of thin
cylindrical conductors connected with each other. At a lateral
coordinate $x$ of a two-dimensional conductor of width $L$, when a
constant current density J is assumed to flow in the z-direction, the
magnetic field along a direction (y-direction) perpendicular to the
surface is calculated as \be H= \frac{2}{c} \int_0^L \frac{d \, X j
(X)}{(X-x)} \, , \lab{spin1} \ee  and then $H= (2j/c) Ln(|x-L|/|x|)$ if
$ j$ is constant. Magnetic field $H$ diverges at the edge, as shown in
Figure 1, and becomes zero at the center of the width of the slab. On the
other hand, for a slab-geometry thin-film conductor with thickness
$d$, the numerically simulated dependence of $H$ as a function of $x$
shows logarithmic divergence of $H$ when we decrease $d$ at
fixed $L$.

\begin{figure}
\begin{center}
\includegraphics[angle=0,width=8cm,scale=1.0]{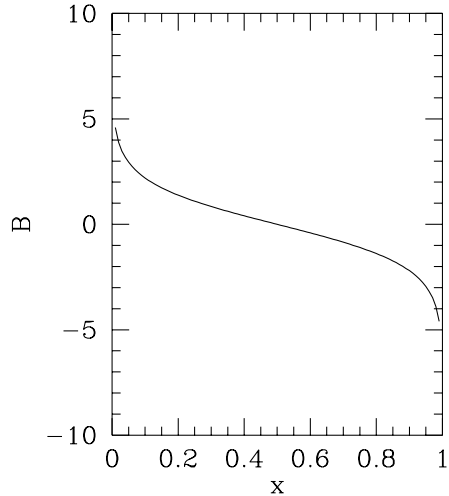}
\end{center}
\caption{ Plot of the dependence of the component of magnetic
induction $10^{-1}B/(\mu_0 \mu)$ (in $A/m$ units) perpendicular to
width $L$ of the sample as a function of observable point coordinate $x$.
Numerical values of magnetic-permeability constants $ \mu_0$ and
$\mu$ are described in the text. } \lab{fig}
\end{figure}

To examine whether, indeed, this CIMF  is responsible for the SHE or
not, we evaluate the strength of magnetic field required for the
spin polarization of the carriers indicated in the experiment and
compare it with that for CIMF.

\section{The gas of electrons in the experiment of Kato {\it et al.} \ci{kato}}

For comparison, we consider the case of the strained semiconductor, because there is
information in this on the density of polarized electrons. The gas
of electrons in the heterostructure $n$-type $In_{0.07}Ga_{0.93}As$
is provided by doped $Si$ atoms with density $3\times 10^{16}$
cm$^{-3}$; therefore, the density of electrons is $\rho=3\times
10^{16}$ cm$^{-3}$. The thickness and width of the sample are
$d=500$ nm and $L=33$ $\mu$m, respectively. The SHE measurement
was carried out at $10\, K<T<60\, K$. The density of polarized
electrons had the systematic error in the interval $+48 \div (-38)$
percent; therefore, we use the ratio of the density of polarized
electrons to the full density of electrons: $10^{-4}$ \ci{kato1}.
The data for the Lande factor  in the bulk of a semiconductor
\ci{kato1}, effective mass \ci{kaestner} and mobility \ci{kikkawa}
of electron carriers are the following: $g=0.64$, $m^*=0.068m$, and
$\mu_e\approx 5400$ cm$^2/(V\cdot s)$ (for mobility, we used the
value for $n$-type $Si$-doped $GaAs$ with an electron density of $
10^{16}$ cm$^{-3}$). The information for experimental effective
mass can be compared with Ref. \ci{lee} for
$In_{0.08}Ga_{0.92}As/GaAs$ and Refs. \ci{noh} for other
semiconductor materials.

We find the following numerical ratios: $\lambda_B/d\approx 2\cdot
10^{-2}$ and $\lambda_B/L=1.351\cdot 10^{-4}$, where
$\lambda_B=\hbar/p_F$ is the de Broglie wavelength with Fermi momentum
$p_F=(3\pi^2 \rho)^{1/3}\hbar$. The calculation of the numerical value
of the Fermi energy ${\cal E_F}/k_B= D\rho^{2/3}/k_B$ \ci{landau1},
where $D=(3\pi^2)^{2/3} \hbar^2/(2 m^*)$ and $k_B$ is the Boltzmann
constant, expressed in Kelvin temperature units ($K$) and measured
from the bottom of the conduction band, gives ${\cal E_F}/k_B= 60.13 \, K$.
These values indicate that the electron gas can be treated as three-dimensional
at the measured temperatures and Fermi-degenerate. The
particles on the Fermi surface are quasi-classical.

To evaluate the strength of magnetic field required for the spin
polarization of electrons in the semiconductor, we use the simple
scheme \ci{vitkalov} of explanation of the effect. There is a Zeeman
splitting  ${\cal E^{\uparrow,\downarrow}}={\cal E_F}\pm g\mu_BH$ of
two subbands of electrons with spin-up $s_z=+1/2=\uparrow$ and
spin-down $s_z=-1/2=\downarrow$ directions of spins in the external
magnetic field. We note that in this definition of directions of
spins, the spin-up component is parallel to the magnetic-field vector
\ci{landau2}. Each energy in ${\cal E^{\uparrow,\downarrow}}$ is
measured from the bottom of its own subband of conductance. In the absence
of a magnetic field, the  numbers of spin-up and spin-down electrons are
equal; therefore, there is no spin polarization. When a magnetic
field is applied, the subband of spin-up electrons shifts down, while
the subband of spin-down electrons goes up. As the Fermi energy is
the same for both spin components of the gas, the electrons with
spin-down spins, whose energy is above the Fermi energy, undergo
spin-flipping and occupy the opened free levels in the subband of
spin-up electrons, below the Fermi energy. The gas has spin-up
polarization along the external magnetic field. In the geometry of
the Kato {\it et al.} experiment, the direction of polarized spins on the edges
coincides with the direction of CIMF, which can be a qualitative
indication that CIMF is responsible for SHE.

From the above explanation of spin polarization, assuming that ${\cal
E^{\uparrow,\downarrow}}=D (\rho^{\uparrow,\downarrow})^{2/3}$, we
obtain the strength of magnetic field: \be H_e= \frac{D}{2 g\mu_B}
((\rho^{\uparrow})^{2/3}-(\rho^{\downarrow})^{2/3}) \, \lab{spin2}
\ee for polarization of $\rho^{\uparrow}-\rho^{\downarrow}$ density
of electrons with spin-up direction of spins. Using the expression
$\mu_B=|e|\hbar/(2mc)$ for the Bohr magneton,
$\rho^{\uparrow}-\rho^{\downarrow}=10^{-4}\rho$,
$\rho^{\uparrow}+\rho^{\downarrow}=\rho$, and substituting the
numerical values for the quantities $gm^*/m$ and elementary flux
quantum of magnetic field $\phi_0=\pi \hbar c/|e|$ in Eq.
\re{spin2}, we find $H_e=56.771\cdot 10^{-4}$ T. This is an
estimate of the strength of magnetic field required for observation of
SHE in the paper of Kato {\it et al.}

For the calculation of the strength of CIMF, we use the expression
\be H\approx \frac{I}{2 \pi L} Ln \left(\frac{L}{d}\right)  \, .
\lab{spin3} \ee Here, $H$ is expressed in SI units (for that, we
replaced the coefficient $4\pi/c \rightarrow 1$) and $I$ is the electric
current flowing along the longitudinal direction of the sample. In the
experiment of Kato {\it et al.}, one gives the electric field $E=25$
mV/$\mu$m instead of current $I$. Employing the relations $j=\sigma
E$ between the density of current $j$ and $E$, and $\sigma=\rho |e|
\mu_e$ between conductivity $\sigma$ and mobility of electrons
$\mu_e$, we find the numerical value for $j$. Then, substituting the
data for $L$ and $d$ for the determination of $I$ through $j$  in
Eq. \re{spin3}, one derives the numerical value $33.526 \, A/m$ for
$H$. The magnetic induction $B_e$ and $H$ are connected with each other
{\it via} the expression $B_e=\mu_0 \mu H$, where $\mu_0= 4\pi \cdot 10^{-7}
\, H/m$ is the magnetic permeability of vacuum and $\mu=1+\chi_P$
with Pauli magnetic susceptibility $\chi_P$. For the three-dimensional
electron gas \ci{landau2} $\chi_P=\mu_B^2 p_F m^*/(\pi^2
\hbar^3)$ and, using the data for density $\rho$ and $m^*$, we obtain
$\chi_P=18.667\cdot 10^{-10}$. Therefore, the numerical value for
strength of CIMF is $B_e=4.213\cdot 10^{-5}$ T. The ratio between
$B_e$ and $H_e$ is $B_e/H_e=7.421\cdot 10^{-3}$, which gives the
estimate $10^{-6}$ of Kato {\it et al.} for the
polarization degree expected from $B_e$. However, on the surface of the semiconductor
$gm^*/m \rightarrow 2$, {\it i.e.}, should be close to the vacuum value, and then
the real quantity is $H_e=1.235\cdot 10^{-4}$ T; hence, $H_e$ has
the same order of magnitude as $B_e$.

\section{The gas of holes in the experiment of  Wunderlich {\it et al.} \ci{wunderlich}}

The experiment was performed on
$(Al,Ga)As$ film doped with acceptor $Be$. The size parameters of
the sample are  $d=1$ nm \ci{kaestner} and $L=1.5$ $\mu$m. Other
quantities describing the experiment are the following: regime for
temperature $T=4.2$ K, effective mass $m^*=0.27m$, current of
holes $I_p=100$ $\mu$A, two-dimensional density of holes $n=
2\cdot 10^{12}$ cm$^{-2}$, $g=0.5$ (this value has been supposed
for $g$, due to its absence in the literature for the investigated or
related materials). As for the Kato {\it et al.} gas of electrons, we assume that
the ratio of density of polarized holes to full density of holes is
$10^{-4}$. The three-dimensional density of holes will be
$\rho_h=n/d$, and we obtain $\rho_h=2\cdot 10^{19} \, cm^{-3}$.

We have $\lambda_B/d \approx 1.19$ and $\lambda_B/L=0.079\cdot
10^{-2}$ for this value of $\rho_h$; therefore, the gas of holes is
two-dimensional. The numerical value of the Fermi energy  ${\cal
E_F}^0/k_B=A n/k_B$, where $A=\pi \hbar^2/ m^*$, of this gas of
holes, measured now from the top of the valence band, yields ${\cal
E_F}^0/k_B= 2.059 \cdot 10^{2} \, K$, which means that at
experimental temperature the gas is Fermi-degenerate.

The Zeeman splitting for the two-dimensional holes is described by
the expression ${\cal E^{\uparrow,\downarrow}}={\cal E_F}^0\mp g\mu_BH$.
Hence, one polarizes the spin-down $s_z=-1/2=\downarrow$ holes
(again, according to the definition, the spin  $s_z=+1/2=\uparrow$ is
parallel to the magnetic-field vector). Assuming ${\cal
E^{\uparrow,\downarrow}}=A n^{\uparrow,\downarrow}$,  we obtain for
the strength of magnetic field the expression \be H_h= \frac{A}{2
g\mu_B} (n^{\downarrow}-n^{\uparrow}) \, \lab{spin4} \ee for
polarization of $n^{\downarrow}-n^{\uparrow}$ density of holes with
spin-down direction of spins. Substituting
$n^{\downarrow}-n^{\uparrow}=10^{-4} n$ and other quantities, as
has been performed above for a three-dimensional electron gas, in Eq.
\re{spin4}, one obtains $H_h=148.15\cdot 10^{-4}$ T. On the other
hand, taking into account that for the present two-dimensional gas
of holes the Pauli susceptibility $\chi_{P,h}=\mu_B^2  m^*/(\pi
\hbar^2 d)$ is $\chi_{P,h}=6.056\cdot 10^{-8}$, the application of Eq.
\re{spin3} with parameters $d=1$ nm, $L=1.5$ $\mu$m and
$I_p=100$ $\mu$A yields $B_h\approx 10^{-4}$ T; therefore,
$B_h/H_h= 0.00676$. On the assumption $gm^*/m \approx 2$ on the
surface of the semiconductor, we derive $H_h= 10^{-3}$ T. $B_h$
should be increased slightly, due to size quantization in the
thickness direction of the sample (for the experimental temperature, one
performs the condition $T \ll \hbar^2/ (m^* d^2 k_B)$ and the gas is in
the ground state of the approximately descriptive one-dimensional
infinite rectangular well, the wave function of which has the  radius
localization  $d/\pi$). Hence, $H_h$ and $B_h$ have the same order
of magnitude.

Finally,  we need to make the following remark. In the calculation of
$B_{e,h}$, it has been supposed that the numerical factor inside of
the logarithmic function in Eq. \re{spin3} is unity. However, one can
show that for a thin slab it tends to $4$. Hence, the numerical value
of $B_{e,h}$ becomes closer to that of $H_{e,h}$. On comparing the
results obtained for both experiments, one can conclude that a possible real
estimate for the strength of magnetic field for the observation of SHE
is $H_{e,h}\approx B_{e,h}\sim 1$ mT. At last, the structure of
CIMF allows us to explain the spatial dependence of spin lifetime
across the sample of Kato {\it et al.}, as long as it is proportional to
the modulus of magnetic field, and to predict the Asymmetric Hall Effect. In
this effect, the two edges of the sample are charged with same-sign charges.

\section{Summary}

We have investigated an estimate for the strength  of
CIMF and shown that it can be close to that required for the observation
of SHE in the experiments of Kato {\it et al.} \ci{kato} and Wunderlich {\it et
al.} \ci{wunderlich}. The reason is that the parameter $gm^*/m$ on
the surface of the semiconductor could have the numerical value for
vacuum. Two  qualitative results obtained might support the fact that
CIMF is responsible for SHE. First, the calculated component of
CIMF, being perpendicular to the main surface of the slab, shows
logarithmic divergence of the ratio of width to thickness of the sample with
opposite signs on the edges, and, second, the direction of polarized
spins in the experiment of Kato {\it et al.} \ci{kato} is along the CIMF, which
is expected for SHE in this magnetic field. From the structure of
CIMF, one could also progress to  the prediction of the Asymmetric Hall
Effect, when both edges of the sample could be  charged with same-sign charges.

\begin{acknowledgments}
 B. A. acknowledges support by Korean
Research Foundation Grant KRF--2004--005--C00044.
\end{acknowledgments}

\newpage

\end{document}